\documentclass[pra,graphicx,reprint,nobibnotes,superscriptaddress,amsmath,showpacs]{revtex4-1}

\usepackage{amsmath}
\usepackage{graphicx}
\usepackage{subfigure}
\usepackage{hyperref}
\begin{document}

\title{Optimal control of high-harmonic generation by intense few-cycle pulses} 

\author{J. Solanp\"a\"a}
\email[]{janne@solanpaa.fi}
\affiliation{Department of Physics, Tampere University of Technology, Tampere FI-33101, Finland}

\author{J. A. Budagosky}
\affiliation{Institute for Biocomputation and Physics of Complex Systems (BIFI) and Zaragoza Center for Advanced Modeling (ZCAM), University of
Zaragoza, E-50009 Zaragoza, Spain}

\author{N. I. Shvetsov-Shilovski}
\affiliation{Department of Physics, Tampere University of Technology, Tampere FI-33101, Finland}

\author{A. Castro}
\affiliation{ARAID Foundation, Mar{\'{\i}}a de Luna 11,  Edificio CEEI Arag{\'{o}}n,
  Zaragoza E-50018, Spain.}
\affiliation{Institute for Biocomputation and Physics of Complex Systems (BIFI) and Zaragoza Center for Advanced Modeling (ZCAM), University of
Zaragoza, E-50009 Zaragoza, Spain}

\author{A. Rubio}
\affiliation{Max Planck Institute for the Structure and Dynamics of Matter, 22761 Hamburg, Germany}
\affiliation{Nano-Bio Spectroscopy Group and ETSF, Departamento de F\'{i}sica de Materiales, CFM CSIC-UPV/EHU-MPC and DIPC, Universidad del Pa\'{i}s Vasco, E-20018 San Sebasti\'{a}n, Spain}

\author{E. R\"as\"anen}
\email[]{esa.rasanen@tut.fi}
\affiliation{Department of Physics, Tampere University of Technology, Tampere FI-33101, Finland}

\date{\today}

\begin{abstract}
At the core of attosecond science lies the ability to generate laser pulses of sub-femtosecond duration.
In tabletop devices the process relies on high-harmonic generation, where a major challenge is
to obtain high yields and high cutoff energies required for the generation of attosecond pulses.
We develop a computational method that can simultaneously resolve these issues by
optimizing the driving pulses using quantum optimal control theory.
Our target functional, an integral over the harmonic yield over a desired energy range,
leads to a remarkable cutoff extension and yield enhancement for a one-dimensional model H-atom.
The physical enhancement process is shown to be twofold: the cutoff extension is of classical origin,
whereas the yield enhancement arises from increased tunneling probability.
The scheme is directly applicable to more realistic models and, within straightforward refinements, also
to experimental verification.
\end{abstract}

\maketitle 


The revolution of attosecond science, i.e., monitoring and controlling
the dynamics of electrons in their native time scale,
relies on the generation of laser pulses with duration of a few
dozens of attoseconds~\cite{[{See, e.g., }]attosecond_pulse_generation}. Such pulses can be
generated by using large-scale free-electron laser facilities~\cite{[{See, e.g., }][{ and references therein.}]fel_review} or
in tabletop devices using high harmonic generation (HHG),
an ultrafast frequency conversion process~\cite{attosecond_pulse_generation}.
Using tabletop devices, however, comes with a price:
the generated attosecond pulses are often too long
and they suffer from low intensity~\cite{attosecond_pulse_generation}.

A high harmonic spectrum has an energy range of nearly constant intensity (plateau),
which ends in a distinctive cutoff~\cite{[{See, e.g., }]hhg_spectrum_form}.
Attosecond pulses are formed from the harmonics on the plateau~\cite{attosecond_pulse_generation}.
Hence, the low amplitude of the pulses is due to low harmonic yield and the pulse duration is determined by the cutoff energy (the higher the energy
the shorter the pulse)~\cite{attosecond_pulse_generation}.
The objectives of increasing
the yield and reducing the pulse duration can be addressed by temporal shaping of the driving pulse --
already experimentally realizable either with multicolor fields or more sophisticated techniques~\cite{[{See, e.g., }]wirth_pulse_shaping,*[{ and }]3genfst}.
Yet a crucial question remains unanswered: how to find the optimal shape of the driving pulse
to enhance HHG?

Numerous previous studies have tackled the issues of cutoff and yield; for a recent review see, e.g.,
Refs.~\cite{review_control_hhg} and~\cite{kohler_hhg_book}.
The main scheme behind the cutoff extension has been using two-color laser
fields~\cite{cutoff_extension_twocolor,cutoff_extension_twocolor2}
or chirped pulses~\cite{chirp_control,chirp_control2,chirp_static},
but also steepening of the carrier wave~\cite{steep_carrier}
or even using a sawtooth pulse should extend the cutoff~\cite{sawtooth}. In addition, also
combined temporal and spatial synthesis of the driving field has been shown to extend the cutoff~\cite{temporal_and_spatial_synthesis}.
A previous study based on quantum optimal control theory (QOCT), for example,
demonstrated some cutoff extension, albeit with a low yield,
by maximizing the ground state occupation at the end of the pulse~\cite{rasanen_optimal_control}.
Yield increase of the plateau has been accomplished, e.g., by
two-color fields~\cite{yield_twocolor_theoretical,yield_twocolor_theoretical2,yield_twocolor,
yield_twocolor_dahlstrom,yield_twocolor_dahlstrom2,yield_twocolor_fleischer}
and also by using a mixture of two target gases~\cite{yield_xuv_assisted}.
In a separate work~\cite{[][{ (submitted)}]selective_theoretical}, some of the authors of the present
work have addressed the selective enhancement of harmonic peaks;
selective harmonic enhancement has been studied using QOCT also in Ref.~\cite{kosloff_pra}, and experimentally, e.g., in Ref.~\cite{selective_tricolor_control}.
Recently also the attosecond pulse generation has been optimized using genetic algorithms~\cite{genetic_optimization}.


In this paper, we provide an efficient computational method to \emph{simultaneously} enhance both
the yield and the cutoff energy of the harmonic plateau
by optimizing the driving pulses with QOCT~\cite{qoct1,qoct2,[{For a recent review on QOCT, see, e.g., }]qoct_review1,*qoct_review2}.
The optimal pulses are found by maximizing the target functional,
an integral over the harmonic yield over a desired energy range.
Surprisingly, the enhancements are achieved with fixed-fluence pulses, i.e.,
the search is performed over the set of pulses with equal duration and fixed fluence (integrated intensity).
We examine in detail the physical origin behind the enhancement, which
is found to be of classical nature to a significant extent.

To demonstrate our method, we use one-dimensional hydrogen with
the soft-Coulomb potential~\cite{soft_coulomb} $V(x)=1/\sqrt{x^2+1}$ as our model system
and the laser-electron interaction
is calculated in the dipole approximation.
The harmonic spectra are calculated from the
Fourier transform of the dipole acceleration $\ddot{d}(\omega)$ as
$S(\omega)=\vert \ddot{d}(\omega)\vert^2 /\omega^2$
as suggested in Ref.~\cite{madsen_hhg_formula}. Unless otherwise
specified, Hartree atomic units (a.u.) are used throughout the paper, i.e., $\hbar=q_e=m_e=1/(4\pi\epsilon_0)=1$.
The time-evolution operator is calculated
using the exponential mid-point rule~\cite{[{See, e.g., }]exp_mid_octopus}
with the Lanczos algorithm~\cite{lanczos_exponential} for the operator exponential; during time-propagation we also use imaginary absorbing boundaries.
We use box size of $4000\ldots 6000$,
grid spacing $0.2\dots 0.3$, and time step of $0.03 \ldots 0.05$; the parameters have been checked
to ensure full convergence. Most of the calculations -- including QOCT discussed below -- are
done in length-gauge using the \texttt{octopus} code~\cite{octopus1,*octopus2}.

In QOCT one solves for a laser pulse $\epsilon(t)$ that maximizes a target functional $J_1[\epsilon]$.
To optimize harmonic spectrum, we have implemented a target of the form
\begin{equation}
\label{eq:j1}
J_1[\epsilon] = \int\limits_{\omega_a}^{\omega_b} \vert \ddot{d}[\epsilon](\omega)\vert^2\,d\omega,
\end{equation}
where $\left[\omega_a,\omega_b\right]$ is the frequency range for
the desired enhancement of the HHG spectrum.
The field $\epsilon$ is represented by a set of parameters,
and maximization of the functional defined in Eq.~(\ref{eq:j1}) amounts to a function
maximization for those parameters. We have used both a gradient-free
algorithm (Newuoa~\cite{newuoa}), and the gradient-based Broyden-Fletcher-Goldfarb-Shannon (BFGS)
algorithm~\cite{bfgs_algorithm} (the expression for the gradient is supplied
by the QOCT).
As we will see, both algorithms provide similar enhancements in the harmonic spectrum.
The optimized pulses are constrained by (i) a finite number of frequencies with the
maximum frequency $\omega_{\max}$, (ii) a fixed pulse length, and (iii)
a fixed fluence which is set to that of a single-frequency \emph{reference} pulse, whose shape will be shown in the figures below.
For each set of pulse constraints, we begin the optimization from several (5-10) random
initial pulses, and report here the best result; it is important to note that QOCT always
converges into a \emph{local} maximum in the parameter space.

\begin{figure}[t]
\centering
\includegraphics[width=\linewidth]{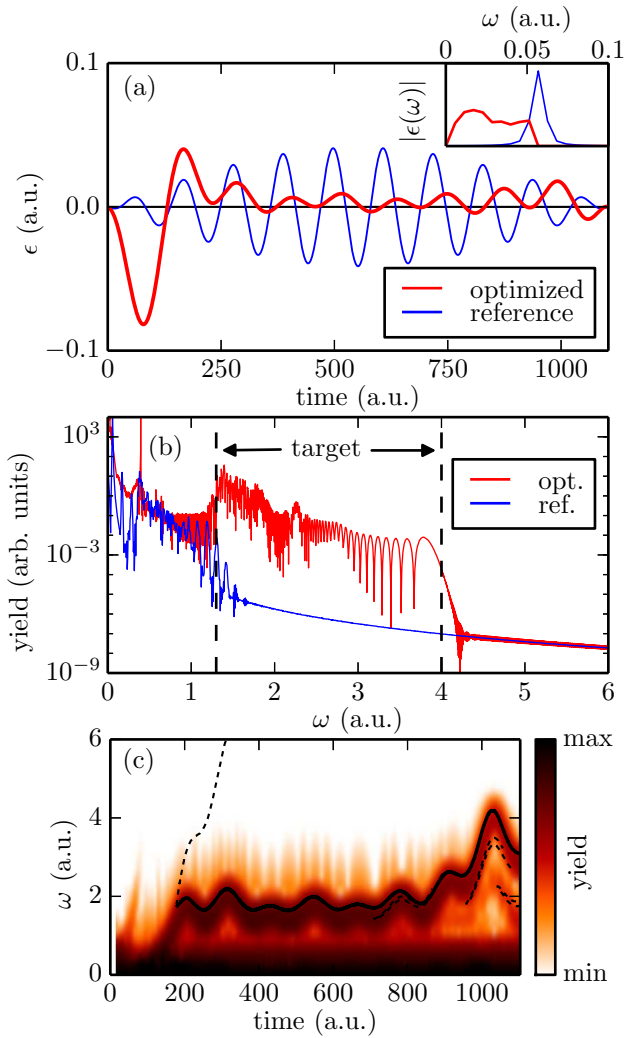}
\caption{Optimization results for the HHG spectrum with the target range $\omega \in \left[1.3,4\right]$ a.u.
The pulse length is $T=1104$ a.u. and the
frequency of the reference pulse is $\omega=0.0569$ a.u.,
equal to the maximum frequency in the optimization. The
fluence is kept constant.
(a) Optimized [red (light gray)] and reference [blue (dark gray)] pulses and their frequencies (inset).
(b) High-harmonic spectra for optimized [red (light gray) and reference [blue (dark gray)] pulses. The target range is shown with vertical dashed lines.
(c) Quantum mechanical time-dependent harmonic spectrum in log-scale [colorscale (grayscale)]
and return energies calculated from the semiclassical model (solid line). Spurious branches from uniform tunneling rate are shown with dashed lines (see text).}
\label{fig1}
\end{figure}

First we apply the Newuoa algorithm to optimize a laser pulse for
HHG in the target interval $\omega \in \left[1.3,4\right]$ a.u. The pulse
length is fixed to $T=1104$ ($26.7$ fs) and the carrier frequency
of the reference pulse is $\omega=0.0569$ a.u.
(wavelength $\lambda \approx 800$ nm corresponding to the typical range of Ti:sapphire lasers),
which we choose to keep as the maximum allowed frequency of the optimized pulse to prevent the formation
of complicated pulses with high frequency components.
The peak intensity
of the reference pulse is $6\times10^{13}\,\text{W}/\text{cm}^2$,
and the fluence is kept constant in the optimization.
The reference and optimized pulses are shown in Fig.~\ref{fig1}(a)
as red (light gray) and blue (dark gray) lines, respectively.
The optimized harmonic spectrum in Fig.~\ref{fig1}(b) completely fulfills
the desired target, and in addition to the cutoff extension,
the yield is also increased by several orders of magnitude.

Next we comment on the two most obvious characteristics of the optimized
pulse in Fig.~\ref{fig1}(a). First, it is important to note that the high-intensity half-cycle
in the beginning is not responsible for the significant increase in the HHG yield and cutoff.
If this part were later in the pulse, the cutoff would be at $\omega\approx2.5$ a.u.
A similar effect is seen if, e.g., the last low-intensity peak is missing.
Secondly, as shown in the inset of Fig.~\ref{fig1}(a), the optimized pulse contains
lower-frequency components.
Indeed, the standard theoretical HHG considerations predict
that lower frequencies should lead to higher cutoff energy due to higher ponderomotive energy.
However, merely using low frequency single-color pulses produces very low yields.
It is the shaped multi-frequency pulses that produce both the large cutoff and high intensities.
Furthermore, in the case of HHG resulting from pulses that have a
single carrier frequency, the harmonic peaks are equally separated by
twice the carrier frequency. In the case of optimized pulses, however,
we find no connection between the frequency components in the pulse
and the HHG peak separations. This is expected in view of the
complexity of the optimized pulse in the time-frequency plane, even
though we applied rather simple pulse constraints as explained above.

The emission process is further demonstrated in \mbox{Fig.~\ref{fig1}(c)}, where the color (grayscale) image shows
the time-frequency map of the quantum dipole acceleration, $\ddot{d}(t,\omega)$. The time-frequency map is calculated as
a discrete short-time Fourier transform~\cite{stft} (STFT) using the Blackman window function~\cite{blackman}.
In essence, the time-axis is split into multiple overlapping windows, and the dipole acceleration is
Fourier-transformed in each window. Finally, we plot the quantity $S(t,\omega)=\vert \ddot{d}(t,\omega)\vert^2/\omega^2$ in log-scale
in analog with the harmonic yield; here $t$ corresponds to the middle of each time-window of the STFTs. $S(t,\omega)$ essentially describes HHG \emph{in time}.
Bicubic interpolation is used for slight visual improvements.
The cutoff extension up to $\omega\lesssim2.5$ a.u. occurs throughout the pulse as it is the effect of the high-intensity peak.
The full extension up to $\omega=4$ a.u., however, occurs only at the end of the pulse. This
clarifies the above-mentioned fact that the complete structure of the optimized pulse is
important.

\begin{figure}
\includegraphics[width=\linewidth]{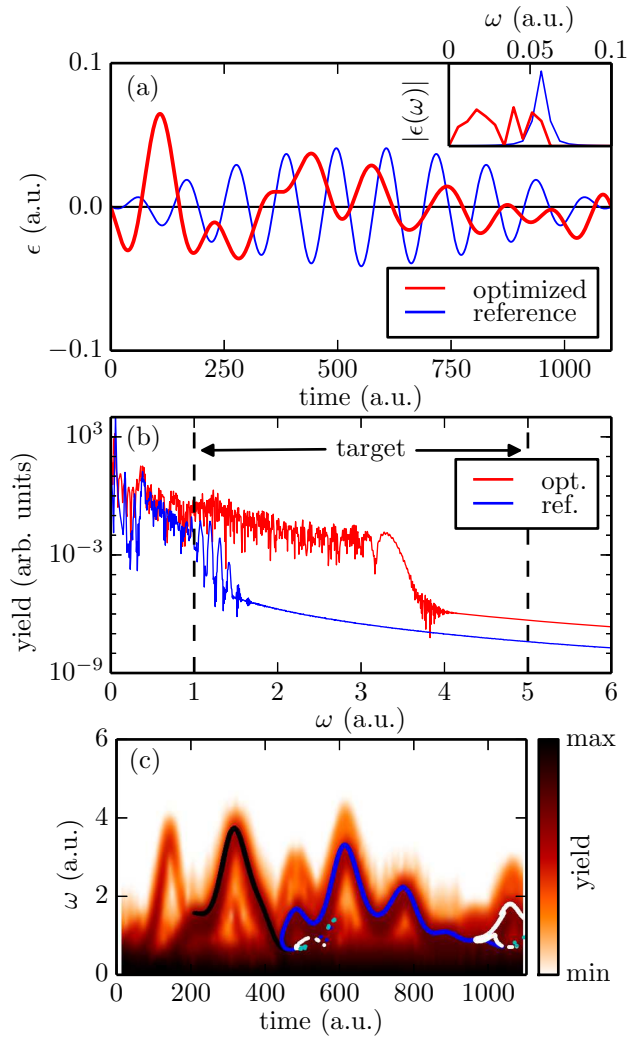}
\caption{Same as Fig.~\ref{fig1} but for an extended target range (up to $\omega=5$ a.u.) and for
the gradient-based BGFS optimization algorithm. In (c), energies of an electron calculated from the semiclassical model upon its first, second, third, and fourth return to the origin are shown with black, blue (medium gray), white, and cyan (light gray) curves, respectively.}
\label{fig2}
\end{figure}

Next we examine the physical origin of the cutoff extension in more detail
by employing semiclassical simulations.
An ensemble of classical trajectories is propagated with initial times $t_0$
distributed according to either a uniform tunneling rate $w(t_0)\sim 1$ or
exponential tunneling rate~\cite{PPT, *[{English: }]PPTen, ADK, *[{English: }]ADKen, KrainovJOSAB}
$w\left(t_0\right)\sim\exp\left\{-\Big[2\left(2I_p\right)^{3/2}\Big]/\Big[3\lvert\epsilon(t_0)\rvert\Big]\right\}$,
where $I_{p}=0.669$ a.u. is the ionization potential of our system.
At the tunnel exit obtained from the classical turning point equation $V\left(x\right)+F_x\left(t\right)x=-I_{p}$,
the velocity is set to zero and the electron is propagated classically.
Upon return of the tunneled electron to the origin, a photon is emitted with frequency corresponding to the kinetic energy of the electron;
also later returns are recorded and taken into account.
Note that in contrast to the three-step (simple man) model~\cite{Corkum93}, where the electron starts from the origin and moves in the laser field only,
the electron in our model starts at the tunnel exit and moves in the combined force field of the laser and the atomic potential.
It should be noted that in contrast to our semiclassical simulation taking the atomic potential into account, the three-step model underestimates the cutoff energy.
For the parameters of Fig.~\ref{fig2} the cutoff calculated from the three-step model corresponds to 3.2 a.u.
(cf. to 4.2 a.u. predicted by semiclassical simulations with binding potential shown in Fig.\ref{fig2}).

The return energy maps of the semiclassical model as a function of the return time (solid curves)
are compared with the time-dependent harmonic spectrum in Fig.~\ref{fig1}(c).
Due to the pulse shape, the electron can return only once to the origin.
With uniform tunneling distribution, the semiclassical model exhibits a few spurious branches (dashed black curves), which are suppressed when using
the exponential tunneling rate. The remarkable agreement between the
semiclassical and quantum descriptions highlights the classical origin of the cutoff extension.

In Fig.~\ref{fig2}(a) we show a BFGS-optimized pulse [red (light gray)] with the same reference pulse [blue (dark gray)] as in Fig.~\ref{fig1}.
The target range is now $\omega \in \left[1,5\right]$ a.u., i.e., considerably larger than in the previous case.
Despite a slightly more complicated temporal shape of the optimized pulse, the resulting HHG spectrum
[Fig.~\ref{fig2}(b)] is similar to the first case. Now, however, the optimized pulse allows
multiple returns of the electron to the origin as shown in Fig.~\ref{fig2}(c) when using an exponential tunneling rate.
Not all of the quantum mechanical harmonic emissions can be found in the semiclassical model with exponential tunneling distribution.
They are, however, allowed by the semiclassical model and visible when using a uniform tunneling rate. Therefore, the semiclassical
picture does agree with the quantum description, but the exponential tunneling distribution does not produce all tunneling
events.

\begin{figure}
\includegraphics[width=\linewidth]{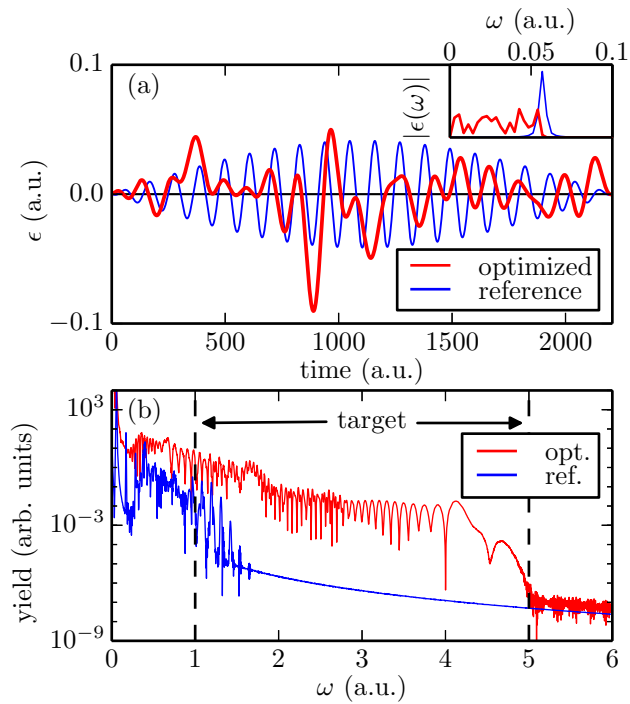}
\caption{Same as Fig.~\ref{fig2}(a-b) but for a longer pulse with $T=2209$~a.u. ($53.5$ fs).
}
\label{fig3}
\end{figure}

Next we double the pulse length while keeping the peak intensity of the reference pulse, the
maximum frequency, and the target HHG range the same (note that the fluence is also doubled).
The BFGS-optimized pulse of Fig.~\ref{fig3}(a) now leads to complete extension
of the cutoff all the way up to $\omega=5$ a.u. as demonstrated
in Fig.~\ref{fig3}(b). This is likely due to higher fluence and more
freedom in the shaping of the longer pulse.

The effect of late returns [see, e.g., Fig.~\ref{fig2}(c)] can be analyzed in the semiclassical picture.
The harmonic spectrum can be calculated as a histogram of the electron energies upon return to the origin
with weights from the exponential tunneling rate (see above).
The resulting spectra demonstrate varying contributions of late returns between different pulses.
Even in the case of pulse of Fig.~\ref{fig2}(a), where late returns are evident,
their contributions to the spectra in the semiclassical models are minimal.
In contrast, for the optimal pulse of Fig.~\ref{fig3},
also the second return plays an important role in enhanced HHG.

The yield increase can be attributed to the increased
tunneling probability compared to the reference pulses.
Indeed, yield increase of comparable, albeit slightly larger, magnitude
can be found when
using single-frequency pulses with the same maximum amplitude
as in the optimized pulses, but the extension of the cutoff does not
reach the optimized results.
Sensitivity of HHG to the pulse amplitude has been previously reported in,
e.g., Refs.~\cite{yield_twocolor} and~\cite{yield_amplitude_dahlstrom}.
The sensitivity is also obvious from the analytic factorization of the HHG rates in Ref.~\cite{hhg_analytic_factorization}.
We emphasize that the yield increase of the presented optimized HHG arises from increased tunneling rate, not
from resonances as, e.g., in Ref.~\cite{yield_twocolor_theoretical}:
in our case a minimum of seven-photon absorption would be required, which is
highly unlikely.

\begin{figure}
\includegraphics[width=\linewidth]{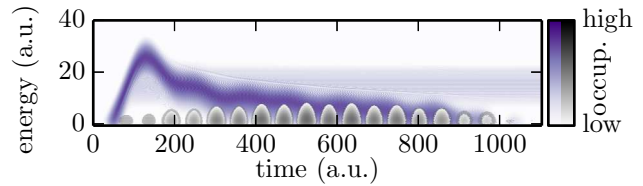}
\caption{Occupations (log-scale) of stationary states in velocity gauge for the reference (gray bulb-shaped structures at the bottom) and optimized (colored structures elsewhere) pulse of Fig.~\ref{fig1}.}
\label{fig4}
\end{figure}

Finally, we verify which stationary states are involved in the
enhanced HHG process. For this purpose,
we solve the time-dependent Schr\"odinger equation in \emph{momentum space
and velocity gauge} by expanding the state in terms of the eigenstates of the field-free Hamiltonian~\cite{nikolay_asymmetry}. Note that the occupations are gauge-dependent.
We find that approximately four lowest bound states are essential for the enhanced HHG, but
ten are required for (nearly) full convergence of the spectrum; the numbers are similar for
reference pulses. However, in the optimized HHG much of the electron density reaches high-energy continuum states, whereas the for the reference pulse the
electron occupation is mostly in the bound states and in the low-energy continuum (see Fig.~\ref{fig4}).

To summarize, we have developed an optimal-control scheme to
simultaneously enhance both the yield and the cutoff energy of high-harmonic generation (HHG).
Our target functional, an integral over the harmonic yield in a desired energy range,
leads to 
significant increase in the HHG yield and cutoff energy
within two different optimization algorithms. Furthermore, we have shown through semiclassical
studies that the extension of the cutoff is of classical origin.
Instead, the increase in the harmonic yield
is found to be due to increased tunneling probability arising from increased peak amplitudes,
while the fluence is kept constant in the optimization.
We note that in higher dimensional models, the harmonic yield will be affected by transversal spreading of the electron wave packet.
However, our preliminary results (not shown here) demonstrate
even the 1D-optimized pulses to provide qualitatively similar
cutoff extension and no significant loss of yield also when applied to a two-dimensional model;
we expect similar tendency also for three dimensions.
In addition, by doing the optimization within the same dimensionality, there can be additional degrees of freedom in the pulse
regarding, e.g., polarization, number of frequency components and pulse sources, which will help counter the issue
of wave packet spreading.

We leave the detailed analysis of realistic pulse constraints
to three-dimensional and many-electron models, where such analysis will be more relevant.
With such refinements, we expect our method
to be usable also in experimental applications, which can have direct
implications in the development of efficient, flexible, and tunable light-emitting tabletop devices.

\begin{acknowledgments}
This work was supported by the Academy of Finland; COST Action CM1204 (XLIC);
the European Community's FP7 through the CRONOS project, grant agreement no. 280879;
the European Research Council Advanced Grant DYNamo (ERC-2010-AdG-267374);
Grupos Consolidados UPV/EHU del Gobierno Vasco (IT578-13); Spanish Grant (FIS2010-21282-C02-01);
and the University of Zaragoza (project UZ2012-CIE-06).
We also acknowledge CSC -- the Finnish IT Center for Science -- for computational resources.
Several Python-extensions~\cite{[{iPython: }]ipython,[{matplotlib: }]matplotlib,
[{Scipy and Numpy: }]scipy,*scipy2} were used for the analysis.
\end{acknowledgments}

\end{document}